\documentclass[aps,prb,longbibliography,twocolumn,superscriptaddress]{revtex4-2}

\usepackage{graphicx}
\usepackage{epstopdf}
\usepackage{miller}
\usepackage{amssymb}
\begin{document}

%\preprint{}

\title{NMR evidence for an antisite-induced magnetic moment on Bi in a topological insulator heterostructure MnBi$_2$Te$_4$/(Bi$_2$Te$_3$)$_n$}


%\author{}
%\thanks{}
%\affiliation{}

\author{R. Kalvig}
\affiliation{Institute of Physics, Polish Academy of Sciences, Aleja Lotnik\'ow 32/46, Warsaw, PL--02668, Poland}
\author{E. Jedryka}
\affiliation{Institute of Physics, Polish Academy of Sciences, Aleja Lotnik\'ow 32/46, Warsaw, PL--02668, Poland}
\author{A. Lynnyk}
\affiliation{Institute of Physics, Polish Academy of Sciences, Aleja Lotnik\'ow 32/46, Warsaw, PL--02668, Poland}
\author{P. Skupi{\'n}ski}
\affiliation{Institute of Physics, Polish Academy of Sciences, Aleja Lotnik\'ow 32/46, Warsaw, PL--02668, Poland}
\author{K. Grasza}
\affiliation{Institute of Physics, Polish Academy of Sciences, Aleja Lotnik\'ow 32/46, Warsaw, PL--02668, Poland}
\author{M. W{\'o}jcik}
\affiliation{Institute of Physics, Polish Academy of Sciences, Aleja Lotnik\'ow 32/46, Warsaw, PL--02668, Poland}


\date{\today}

\begin{abstract}
MnBi$_2$Te$_4$  (MBT) is the first intrinsic magnetic topological insulator, combining a topologically protected surface metallic state and intrinsic magnetic order. A structural compatibility with the nonmagnetic Bi$_2$Te$_3$ (BT) parent compound gives a possibility to create MBT/BT heterostructures and manipulate their magnetic state in view of optimizing the Quantum Anomalous Hall Effect (QAHE). In this work an extensive Nuclear Magnetic Resonance (NMR) study, supported by the bulk magnetization measurements has been performed at 4.2 K on a self-organized single crystal MnBi$_2$Te$_4$/(Bi$_2$Te$_3$)$_n$ heterostructure, obtained from the Mn$_{0.81}$Bi$_{2.06}$Te$_{4.13}$ melt.  $^{55}$Mn and $^{209}$Bi NMR signals have been recorded as a function of the out-of-plane magnetic field  up to 6 T, covering a spin-flop transition (SFT) from the antiferromagnetic (AFM) to the canted antiferromagnetic (CAFM) configuration of the Mn layers. The canting angle at different external field values has been estimated based on NMR data. Presence of the AFM-coupled Mn antisites has been evidenced and shown to induce an antiparallel magnetic moment on Bi atoms within the host Bi  layer. Detection of the induced magnetic moment on bismuth which contributes a new ferromagnetic (FM) component is of utmost importance for understanding the magnetic interactions in the MBT/BT system. These findings have potentially important implications for engineering the QAHE devices.
\end{abstract}

%\keywords{}

\maketitle

\section{Introduction}

Magnetically ordered topological insulators have attracted a great deal of attention during last decade. A combination of topologically protected surface metallic state and intrinsic magnetic order provides excellent ground for investigating Axion Insulating States and Quantum Anomalous Hall Effect (QAHE) \cite{Mogi17,Haldane88}. The possibility of realizing QAHE in zero external magnetic field is particularly sought after as a bridge to develop low power consumption materials and devices \cite{Hirahara17,Gong17,Huang17}. Following  several theoretical predictions and first experimental confirmation it has been demonstrated that that MnBi$_2$Te$_4$ (MBT) -- the first intrinsic magnetic topological insulator -- is a great platform to investigate those elusive physical phenomena \cite{Aliev19,Otrokov19,Otrokov19PRL, Gong19,Li19,Li20,He20,Akhgar22,Gao21}. 

MnBi$_2$Te$_4$ crystalizes in a rhombohedral layered structure with the space group $R\bar3m$ and has the following lattice parameters $a=4.332$ $\AA$ and $c=42.979$ $\AA$ \cite{Lee13}. It is a derivative of Bi$_2$Te$_3$ (BT) – an intensely studied  non-magnetic topological insulator.  In MBT the hexagonally arranged monoatomic layers are stacked along the c-direction [0001] in the sequence Te – Bi – Te – Mn – Te – Bi – Te, forming a septuple layer (SL) as shown in Fig. 1(a). The subsequent septuple layers are weakly bonded by van der Waals forces, giving a possibility of interlacing them with several layers of an isostructural BT.  Such heterostructures  - denoted as  MnBi$_2$Te$_4$/(Bi$_2$Te$_3$)$_n$  exhibit a variety of the long-range magnetic ordering, creating a tunable platform of different topological  states of matter \cite{Klimovskikh20,Wu20}.  They consist of the magnetic septuplets alternating with non-magnetic quintuple layers (QL) of Bi$_2$Te$_3$ (Te – Bi – Te – Bi –Te). The $n$ number in the formula denotes a number of the QLs separating the subsequent SLs. 

Structurally perfect MnBi$_2$Te$_4$ ($n=0$) is an A-type antiferromagnet (AFM) along the c-axis [0001] with a transition temperature T$_{N}$ = 24 K \cite{Otrokov19,Chen19PRM}. Within each septuplet the Mn atoms are ferromagnetically (FM) ordered along the easy c-axis and are AFM coupled to the Mn layer in the neighbor SL. The AFM coupling persist also when the subsequent SLs are separated by one or two QLs, whereas for $n\geq3$ the AFM interaction is not active anymore \cite{Klimovskikh20}. The AFM coupling can also be broken by application of the external magnetic field along the [0001] direction: a spin-flop transition (SFT) has been observed between 3.4 T and 3.57 T \cite{Otrokov19,Yan19,Li20,Lee19} at 2 K leading to a metastable canted-AFM (CAFM) state. It has recently been shown that a self-organized growth is a preferable way to achieve the MnBi$_2$Te$_4$/(Bi$_2$Te$_3$)$_n$  heterostructures, giving materials that are magnetically better organized than the substitutional Mn-doping of the BT crystals \cite{Sitnicka22}. This preparation method has an advantage of achieving a considerable volume of a material in form of a single crystal, but does not allow to control precisely the distribution of the QLs, so the $n$ number reveals a certain distribution over the sample volume. Other potential structural defects include a magnetic disorder within SLs  in form of Mn$_{Bi}$ and Bi$_{Mn}$ antisites as well as Mn/Bi intermixing within QLs \cite{Sitnicka22}. The structural defects lead to the onset of additional magnetic interactions, and have a deteriorating effect on the QAHE state \cite{Wu20,Huang20,Islam23}, stressing the importance of understanding the role they play in the system.

\begin{figure*}[ht]
\includegraphics[width=\textwidth, scale=0.33]{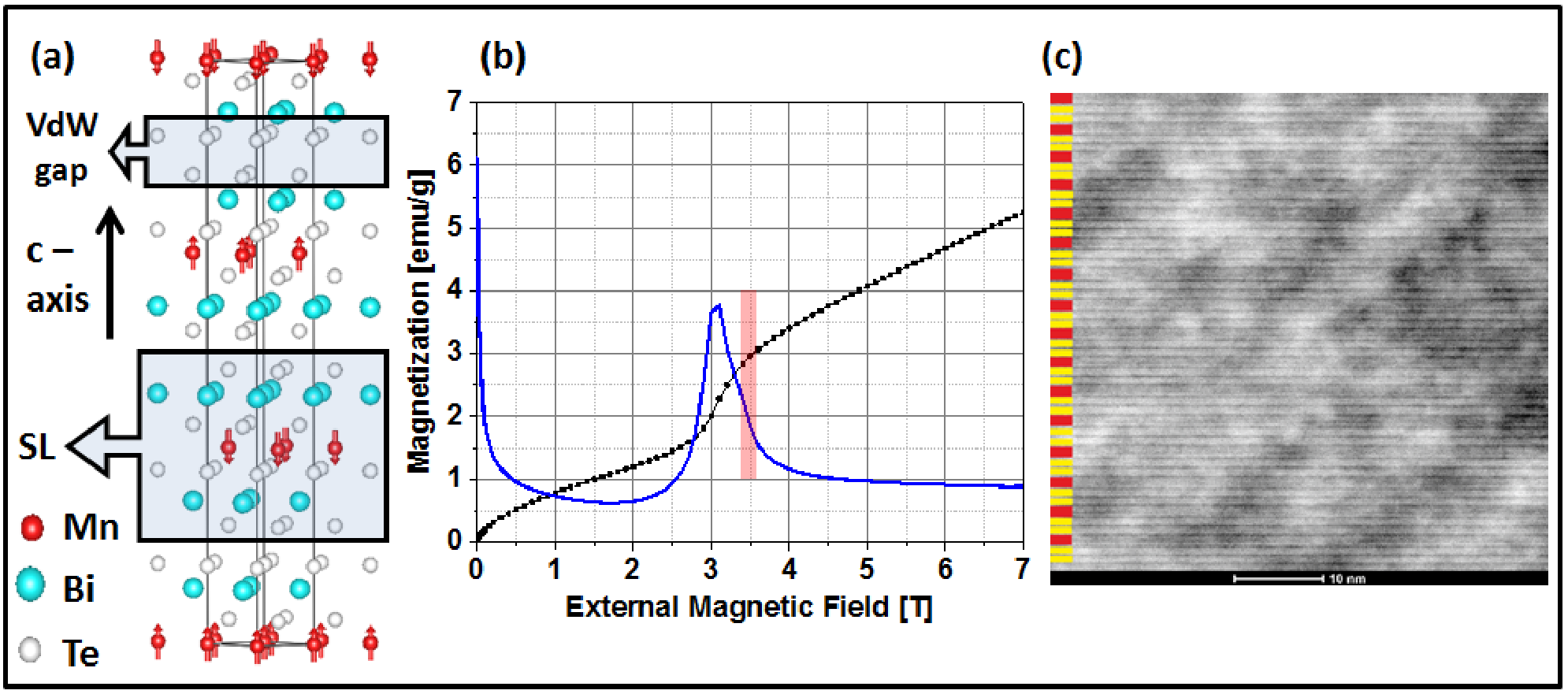}
\caption{\footnotesize (a) Schematic magnetic structure of the MnBi$_2$Te$_4$. Mn atoms self-organize as a central layer within a septuplet (Te -- Bi -- Te -- Mn -- Te -- Bi -- Te) and reveal FM order. Magnetic moments (represented here by red vectors) are AFM coupled to the subsequent Mn-planes.  Bi and Te atoms are represented with with full cyan and white open circles, respectively. This drawing was prepared by means of VESTA software \cite{Momma11}.
 (b) Experimental magnetization curve (black) and its first derivative (blue) recorded from an inhomogeneous MnBi$_2$Te$_4$/(Bi$_2$Te$_3$)$_n$ heterostructure measured at 4.2 K with the external magnetic field applied along [0001] direction. (c) TEM image of the studied heterostructure of MnBi$_2$Te$_4$/(Bi$_2$Te$_3$)$_n$. The selected fragment shows a part of the sample where the number of QLs (highlighted in the margin in yellow) separating the SLs (highlighted in the margin in red) is predominantly $n=2$.}
\label{fig_1}
\end{figure*}

Nuclear Magnetic Resonance (NMR) experiment is an invaluable tool to disentangle the magnetic response and identify different magnetic phases present in a multiphase magnetic system, since it is a site-sensitive  and element-sensitive probe. The experiment involves  the entire volume of the material providing a distribution of intrinsic local magnetic fields that depend on a magnetic state of a probed atom and its interaction with the magnetic environment. Previously reported  NMR studies in the MBT system involved polycrystalline samples of MnBi$_2$Te$_4$/(Bi$_2$Te$_3$)$_n$ with ($n=0, 1, 2$) measured in the external magnetic fields up to 3 T \cite{Sahoo24}. In this work we investigate the local structural and magnetic properties of a self-organized MnBi$_2$Te$_4$/(Bi$_2$Te$_3$)$_n$ single crystal, where $n$ is the number of QLs separating SLs which in our case is $n=2$ on average as determined from the Transmission Electron Microscopy (TEM) image (Fig. 1 (c)). The NMR spectra have been recorded at different values of the external magnetic field up to 6 T i.e. beyond the MBT spin flop field and analyzed in the context of bulk magnetization measurements. A well-defined orientation of the external magnetic field with respect to the crystal axes made it possible to follow the SFT to the CAFM state, determine the canting angle $\theta$ and investigate the orientation of the manganese antisite magnetic moments. Most importantly, this study provides the first direct experimental evidence for an antisite -- induced magnetic moment on Bi atoms.

\section{Experimental details}

The sample under study was a single crystal obtained by the Bridgman growth process from the Mn$_{0.81}$Bi$_{2.06}$Te$_{4.13}$ melt and no chemical inhomogeneity has been observed by means of energy dispersive x-ray analysis. The details of sample preparation and initial characterization have been described in Ref. \cite{Sitnicka22}. NMR experiments were performed using an automatic, phase-sensitive spin-echo spectrometer with a tunable probe \cite{Nadolski95}. $^{209}$Bi and $^{55}$Mn resonances have been investigated at 4.2 K by sweeping frequency in the range 60 – 200 MHz and 350 – 440 MHz, with a step of 1 MHz. The external magnetic field was applied along the c-axis (i.e. [0001] direction) of the studied crystal and varied in the range 0-6 T. Experimental NMR spectra have been corrected for the NMR enhancement factor using the Panissod protocol \cite{Panissod00}. Macroscopic magnetic characterization was carried out with a Quantum Design superconducting quantum interference device magnetometer (SQUID) at a constant temperature of 4 K. 

\section{Results and Discussion}

\subsection{Macroscopic magnetization characterization}

The magnetic structure expected for the perfectly ordered MnBi$_2$Te$_4$/(Bi$_2$Te$_3$)$_n$  system \cite{Otrokov19,Gong19} is presented in the Fig. 1(a), whereas Fig. 1(b) presents the results of macroscopic magnetization measurements performed on the studied sample vs. the external magnetic field up to 7 T applied along the [0001] direction (i.e. perpendicular to the Mn planes).

Unlike the M(H) curves reported for the MBT single crystal ($n=0$) \cite{Yan19} revealing at 2 K a close to zero magnetization below the SFT, we observe at low fields an uncompensated FM contribution. This can be easily linked to a different  morphology of our sample. The sample studied in Ref. \cite{Yan19} was carefully processed to avoid the presence of interlacing BT quintuple layers -- zero magnetization below the SFT indicates a fully compensated AFM alignment and confirms a perfect structure of that material. In contrast to that, our self-organized sample reveals a superposition of phases with different n values ($n=0, 1, 2, 3$ etc.) within the sample volume. It has been shown \cite{Klimovskikh20} that the strength of the AFM coupling between the neighboring Mn layers rapidly drops for $n=1$ and even more for $n=2$, whereas for $n=3$ it practically disappears, leading to a FM behavior. Another likely source of FM in our sample is the presence of antisite defects contributing a non-zero magnetic moment below the SFT. The $Mn_{Bi}$ antisites consisting of Mn atoms substituting Bi in QLs were shown to couple FM to the Mn layers within SLs \cite{Sitnicka22,Liu21,Yan22}. The presence of a FM component attributed to structural disorder has been also reported in several other MnBi$_2$Te$_4$/(Bi$_2$Te$_3$)$_n$ systems \cite{Huang20,Liu21,Sitnicka22,Islam23,Sahoo24}. A similar M(H) curve was reported e.g. for epitaxial MnBi$_2$Te$_4$/GaAs(111) films grown by molecular beam epitaxy \cite{Bac22}. 
A pink highlighted rectangle in Fig. 1(b) indicates the field range (3.4 T – 3.57 T) where the M(H) curves from the stoichiometric MBT samples ($n=0$) were reported to reveal a characteristic sharp step indicating a SFT from a perfect AFM alignment to a CAFM state. This feature has been observed in MBT single crystals at 3.5 T\cite{Otrokov19,Yan19} and in powders at 3.4 T\cite{Li20}. It has been shown that the SFT field is dramatically reduced in case where the neighbor SLs are separated by BT quintuplets (i.e. in samples with $n\neq 0$). Already for $n=1$ the M(H) loops reported at 2 K indicate the spin flop field as low as 0.2 T \cite{Klimovskikh20,Hu20,Wu19}. Therefore the M(H) curve shown in the Fig. 1(b) revealing a characteristic inflection in vicinity of 3 T (better visible on the first derivative curve) can be considered as an indicator of a significant number of purely MBT fragments present in our sample, alongside with the FM phase discussed above. The SFT in our sample takes place over a rather extended field range due to a size distribution of pure MBT fragments alternating with FM--coupled segments. Above 3.5 T, a further linear increase of magnetization is observed, related to the rotation of canted Mn spins towards the c-axis. The magnetic saturation is not reached within the available magnetic field range (0–7 T), in line with the previously reported data showing that the MBT system requires around 60 T to reach magnetic saturation \cite{Lai21}. This macroscopic magnetic characterization will serve as a basis for understanding the details of the NMR results.  

\subsection{$^{55}$Mn NMR}

Figure 2(a) presents the $^{55}$Mn NMR spectra recorded at 4.2 K at different values of the external magnetic field applied along the [0001] direction, i.e., along the crystallographic c-axis. First, we note the presence of two distinct spectrum components (here marked as red and black, respectively), their frequency shifting in opposite directions with increasing magnetic field strength.

\begin{figure}[ht]
\includegraphics[width=\columnwidth]{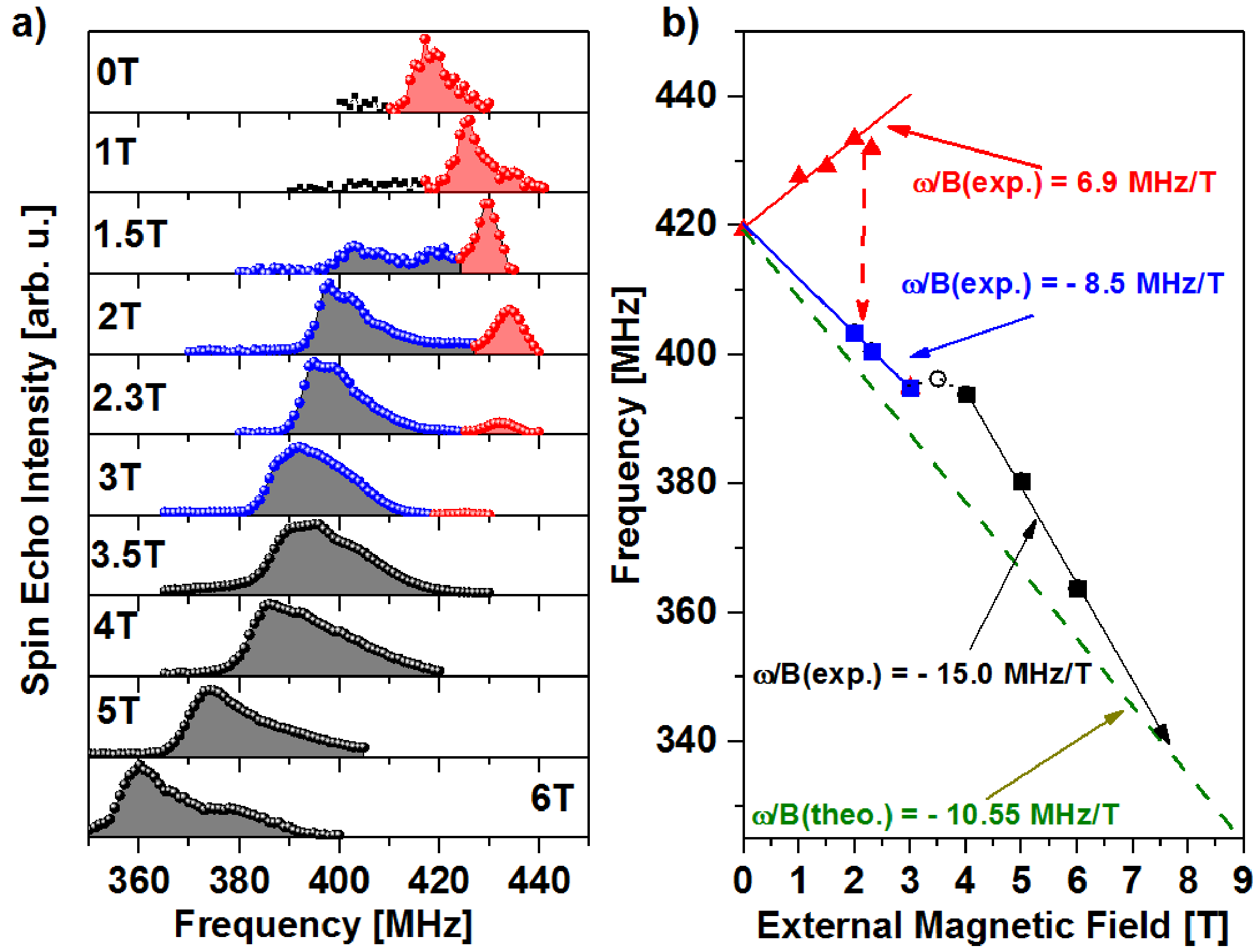}
\caption{\footnotesize (a) $^{55}$Mn NMR spectra recorded at 4.2 K from an inhomogeneous MnBi$_2$Te$_4$/(Bi$_2$Te$_3$)$_n$ heterostructure at different values of external magnetic field applied along the crystallographic \textit{c} axis. (b) $^{55}$Mn NMR frequency $\omega$ (spectrum gravity center) versus the external magnetic field $\vec{B}_{ext}$. Red/blue/black symbols denote the respective NMR signals from the Mn spins oriented  antiparallel/almost parallel/canted with respect to the external field. The dashed green line describes the hypothetical $\omega(\vec{B}_{ext})$ dependence in case of perfectly parallel alignment.}
\label{fig_2}
\end{figure}

To understand the origin of these two signals we recall the basic NMR equation, determining the NMR frequency  $\omega$.
\begin{equation}
 \omega=\gamma|\vec{B}_{eff}| \simeq \gamma|\vec{B}_{hf}+\vec{B}_{ext}|      
\end{equation}

where $\gamma$ is the gyromagnetic ratio, characteristic for a specific nucleus and $\vec{B}_{eff}$ denotes the strength of the effective magnetic field at the nucleus site, which in the absence of an external magnetic field is practically equal to the hyperfine field $\vec{B}_{hf}$. 

The main contribution to hyperfine field comes from the contact Fermi term $\vec{B}_{hf}^{cf}$, which is due to the electronic magnetic moment of the same atom  and directed antiparallel to this moment. Therefore the respective upshift or downshift of the NMR frequency in presence of the external field is a fingerprint of antiparallel/parallel alignment of the magnetic moment vs. the field direction.
In addition, magnetic moments of the surrounding atoms may also contribute to a hyperfine field, providing a transferred hyperfine field $\vec{B}_{hf}^{cf,trans}$ \cite{Jedryka21}:

\begin{equation}
 \vec{B}_{hf}^{cf}=\vec{B}_{hf}^{cf,core}+\vec{B}_{hf}^{cf,trans}=A\vec{\mu}_{on-site}+A'\Sigma\vec{\mu}_{s}
\end{equation}

where the A and $A'$ denote the respective hyperfine interaction constants. 

\begin{figure}[ht]
\includegraphics[width=\columnwidth]{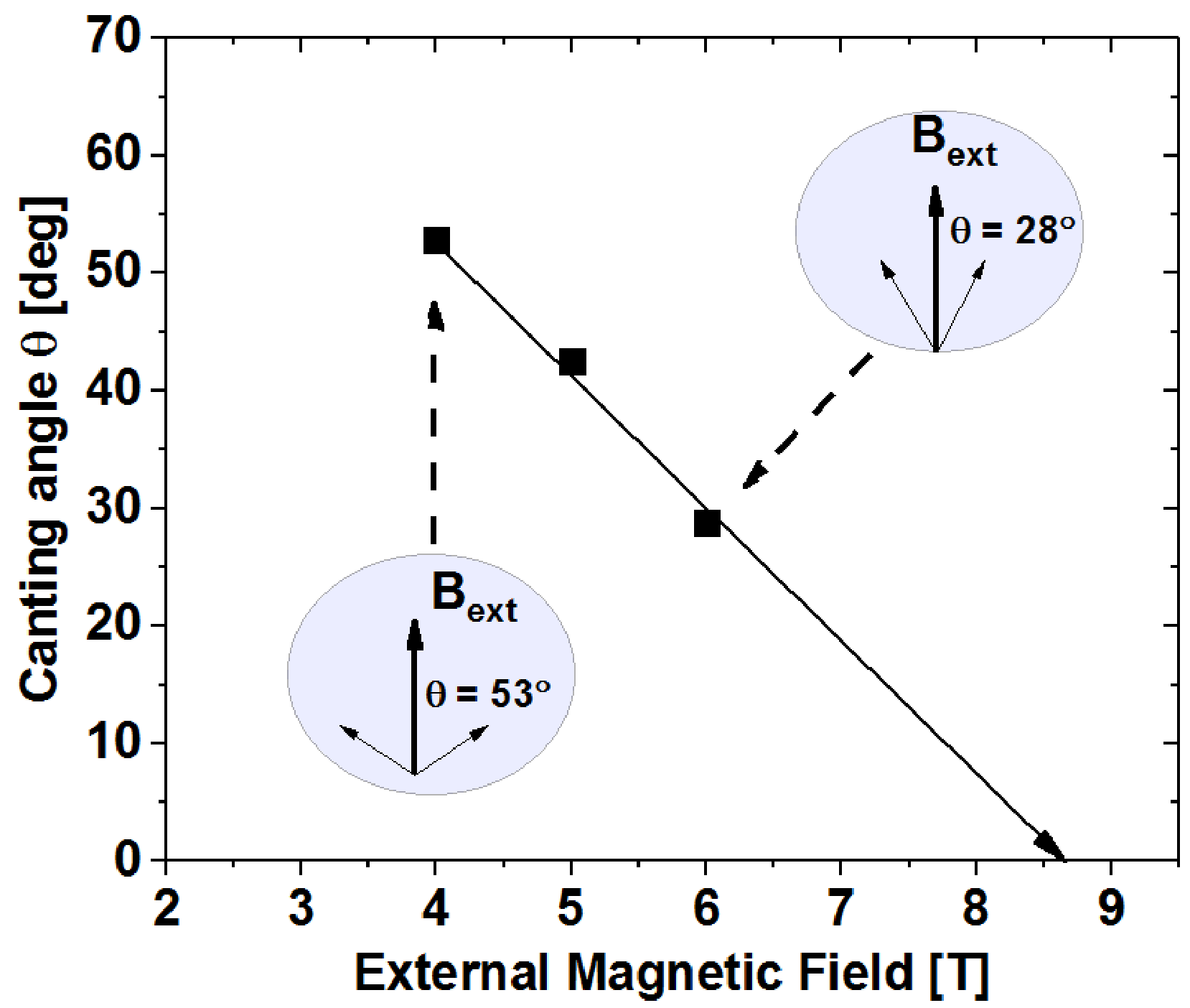}
\caption{\footnotesize Canting angle $\theta$ dependence on the external field applied along the c-axis. Data points computed using eqn. 3.}
\label{fig_3}
\end{figure}

In view of the above, considering the morphology of our sample, we can readily attribute the main “black” $^{55}$Mn NMR signal to a FM component of the material, whereas the “red” $^{55}$Mn NMR signal is a fingerprint of the Mn magnetic moments that are clearly oriented opposite to the external field lower than 1.5 T, i.e. the external field is oriented along their hyperfine field, leading to the frequency upshift.
Let us first consider the NMR signals contributing to the “black” part of the spectrum. Contrary to the expectations,  the substantial FM component which is clearly evidenced at low fields by the magnetization measurements, does not give a measurable NMR signal below 1 T,  most probably due to a fast relaxation, experimentally estimated at $T_2=8 \mu$s. The well-resolved NMR spectrum could be recorded only at and above 1.5 T, increasing in intensity up to 3.5 T and downshifting in frequency over the entire investigated field range. The frequency position of the spectrum gravity center is plotted in Fig. 2(b) as a function of external magnetic field. Interestingly, over the field range between 3 and 3.5 T (corresponding to the SFT) the slope of $\omega/B_{ext}$ dramatically changes from 8.5 MHz/T to 15 MHz/T, indicating two different contributions to the overall NMR signal. Let us recall that the hyperfine field is antiparallel to the electronic magnetic moment, therefore in FM configuration the eqn. 1 takes the form
$\omega\simeq\gamma(B_{hf}-B_{ext})$. Obviously, in case of perfect alignment between the external field and the hyperfine field on nucleus this slope would be equal to the gyromagnetic ratio $\gamma_{Mn}=10.55$ MHz/T. Indeed, previous NMR studies have shown that in samples with a well-defined stoichiometry ($n=1$ and $n=2$) the slope of $\omega/B_{ext}$ line is very close to $\gamma_{Mn}$ since the SFT for such samples takes place at very low fields and the Mn magnetic moments are practically aligned along the external field \cite{Sahoo24}. This is clearly not the case of our complex heterostructure. The slope of 8.5 MHz/T means that the Mn moments contributing to the NMR signal in the field range 1-3 T do not readily follow the external magnetic field. This is due to the diversity of magnetic interactions in that field range, as described in subsection A. 

Between 3 T and 3.5 T a spin-flop starts and the mean NMR frequency is almost constant in this transition region, indicating that the overall NMR signal is now predominantly due to a new magnetic component, overwhelming the weak contribution recorded below 3 T (blue symbols in the Fig. 2(b)).
Above 3.5 T the two manganese sublattices within the pure MBT regions ($n=0$) that were previously AFM--coupled and thus invisible in NMR, switch their orientation to a canted state (CAFM) - the magnetic moments become tilted by an angle $\theta$ from the FM axis. Consequently, the external field component contributing to the effective field on Mn nuclei is now equal to $B_{ext}cos\theta$ and the eqn. 1 takes the form $\omega\simeq\gamma(B_{hf}-B_{ext}cos\theta)$.  Above 3.5 T the $\omega/B_{ext}$ slope shifts to 15 MHz/T exceeding the value of $\gamma_{Mn}$ and thus showing an increasing negative contribution of the $B_{ext}cos\theta$. The NMR frequency is determined by the modified eqn. 1 : 

\begin{eqnarray}
\omega=\gamma|\vec{B}_{eff}|=\gamma_{Mn}(B_{hf}-B_{ext}cos\theta)=\nonumber \\
=420-\gamma_{Mn}B_{ext}cos\theta
\end{eqnarray}

where 420 MHz is the NMR frequency due to the intrinsic hyperfine field determined from the extrapolation to zero external field whereas $\gamma_{Mn}cos\theta = 15$ MHz/T is the “effective” gyromagnetic ratio in the canted state. 

Fig. 3 shows the $\theta(B_{ext})$ dependence calculated from the eqn. 3. Bearing in mind that the data points in Fig. 2(b) represent the gravity center of rather broad spectra, the obtained values of $\theta$ angle must be regarded as an estimate and not an absolute value. They do give, however, an insight on the variation of the canting angle as a function of the external field. As seen in the Fig. 3, the extrapolation to $\theta = 0$ gives the value of around 8.5 T, which is in good agreement with the reported high field magnetization measurements in MBT \cite{Lai21}.  

On the other hand, the upshifting “red” $^{55}$Mn NMR signal is a fingerprint of some Mn magnetic moments that are clearly oriented opposite to the external field lower than 1.5 T, so the external field acts along the hyperfine field and the NMR frequency increases. These are the Mn$_{Bi}$ antisites within the SLs -- their antiparallel alignment with respect to the Mn planes has been already demonstrated  in MBT by high-field magnetization measurements \cite{Lai21} and confirmed by NMR studies in the polycrystalline MnBi$_2$Te$_4$/(Bi$_2$Te$_3$)$_n$  system \cite{Sahoo24}.  We note, however, that the observed frequency of the antisite $^{55}$Mn NMR signal  (around 419 MHz) differs from that reported in Ref. \cite{Sahoo24} (around 470 MHz), which is likely due to a different form of the studied samples (our single crystal vs. polycrystalline samples studied in Ref. \cite{Sahoo24}). The individual antisite Mn moments are susceptible to the r.f. excitation and give the NMR signal even at zero field. As the external field gets stronger, they align with the field adding to the magnetization of the uncoupled Mn layers, and contribute to the “black” spectrum component. Therefore the “red” signal eventually disappears.

\subsection{$^{209}$Bi NMR}

\begin{figure}
\includegraphics[width=\columnwidth]{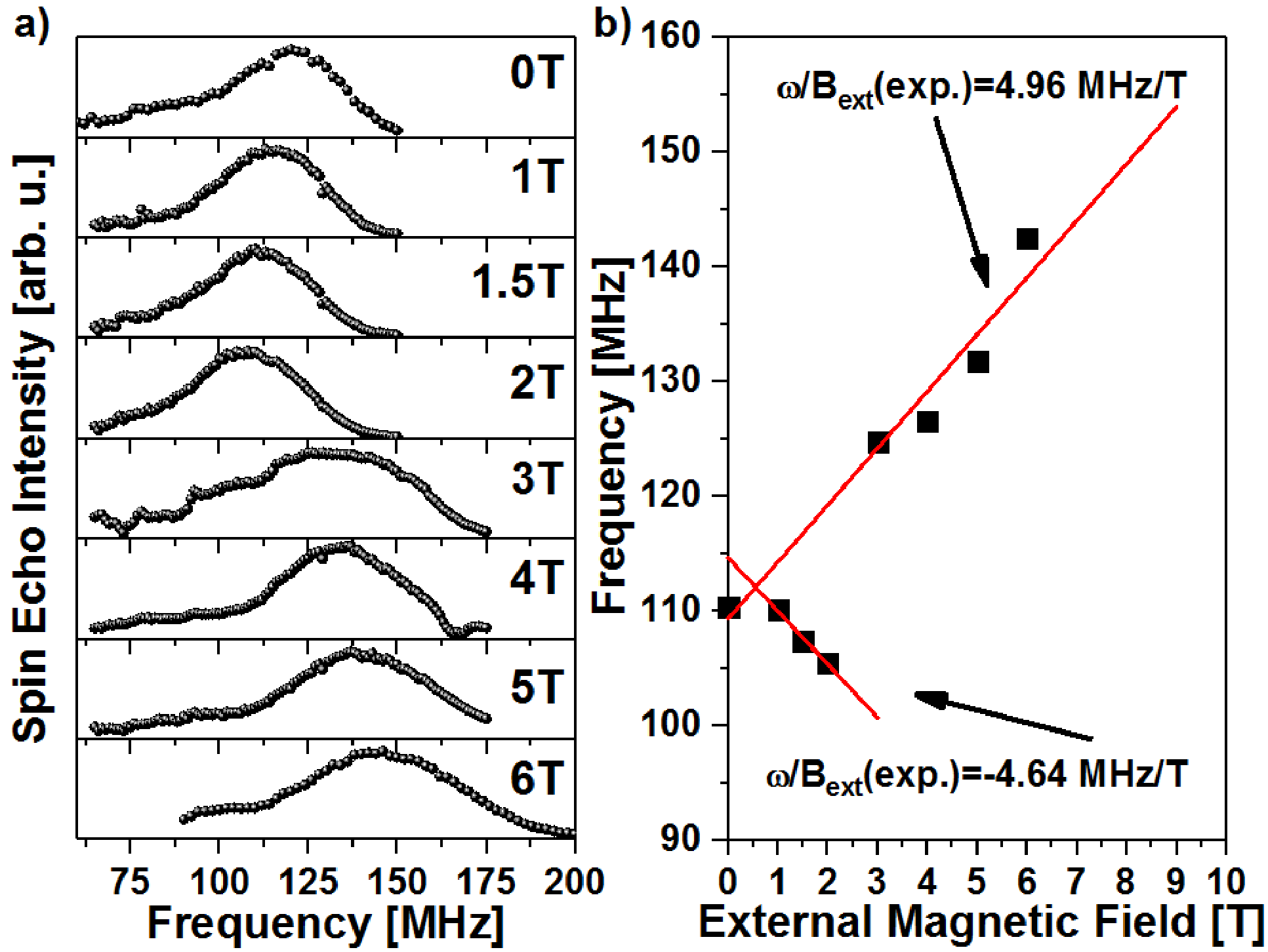}
\caption{\footnotesize (a) $^{209}$Bi NMR spectra recorded at 4.2 K from an inhomogeneous MnBi$_2$Te$_4$/(Bi$_2$Te$_3$)$_n$ heterostructure.  The external magnetic field was applied perpendicular to the film plane (i.e., along the crystallographic \textit{c} axis) (b) $^{209}$Bi NMR frequency versus magnetic field applied in the \textit{c} direction of MnBi$_2$Te$_4$.}
\label{fig_4}
\end{figure}

In addition to the $^{55}$Mn NMR  signals  already discussed, at zero field we observe another strong NMR signal around  110 MHz.  Whereas Mn is the only magnetic element in the system, the magnetic interactions can induce a magnetic moment on the otherwise non-magnetic elements, in this case Bi or Te.  Considering the difference in the natural abundance  of the respective isotopes $N=100\%$ for  $^{209}$Bi (spin $I=9/2$) vs. $N=7\%$ for $^{125}$Te (spin $I=1/2$) and the fact that the NMR signal intensity (expressed as $N(I+1)I$) of bismuth is two orders of magnitude higher than that of Te, the signal observed around 110 MHz can be  undoubtedly attributed to  $^{209}$Bi nuclei. A very large linewidth of a signal around 110 MHz can be readily explained by inhomogeneity of quadrupolar interactions, which would be absent in case of Te (spin $I=1/2$) whereas expected strong at Bi (spin $I=9/2$). Taking into account the gyromagnetic ratio  $\gamma_{Bi}  = 6.84$ MHz/T  we obtain that the effective field on  $^{209}$Bi amounts to around 16 T. Such a strong field implies the presence of a magnetic moment on Bi, generated by a hybridization of manganese 3d orbitals and bismuth valence 6p, 6s orbitals. Interestingly, the presence of a magnetic moment on Se in the isostructural heterostructures MnBi$_2$Se$_4$/(Bi$_2$Se$_3$)$_n$ has recently been suggested based on the x-ray magnetic circular dichroism study \cite{Fukushima24}. To the best of our knowledge, our observation of the  $^{209}$Bi NMR signal in the present study provides the first direct evidence that bismuth also reveals an induced magnetic moment in the MnBi$_2$Te$_4$/(Bi$_2$Te$_3$)$_n$  heterostructures. 

We have further investigated this effect  by recording the NMR spectra at different values of the external magnetic field applied along the [0001] direction (Fig. 4 (a)).  The frequency position of the $^{209}$Bi NMR signal,  plotted as a function of external magnetic field in the Fig. 4 (b), reveals above 2 T a linear dependence with the positive slope of 4.96 MHz/T. This is close to the $^{209}$Bi gyromagnetic ratio $\gamma_{Bi}  = 6.84$ MHz/T, giving an additional proof for the correctness of our assignment of this signal to $^{209}$Bi, rather than $^{125}$Te ($\gamma_{Te}  = 13.45$ MHz/T). 

Interestingly, the $^{209}$Bi NMR  spectra reveal a field dependence closely related to that observed for the antisite Mn$_{Bi}$ signal. Just like in the case of a “red” $^{55}$Mn NMR spectrum we can distinguish two field regimes: below 2 T the NMR frequency shifts towards lower frequencies while above 2 T a steady upshift is observed. This means that the induced magnetic moment on Bi is closely linked to manganese in the antisite positions and oriented antiparallel to the parent Mn$_{Bi}$ moment,  located in the same atomic layer. Mn$_{Bi}$ moment polarizes six bismuth nearest neighbors and gives rise to the observed strong $^{209}$Bi NMR signal. It must be stressed that the straightforward  attribution of the origin of Bi moment to the interaction with the Mn$_{Bi}$  antisites could only be possible based on the evolution of the $^{55}$Mn and $^{209}$Bi NMR spectra as a function of magnetic field with the well-defined, out-of-plane orientation.

\section{Conclusions}

NMR experiments and magnetization measurements performed in this study demonstrate that the complex MnBi$_2$Te$_4$/(Bi$_2$Te$_3$)$_n$ heterostructure obtained via self-organized growth from the melt contains a significant amount of pure MBT environments. A SFT from an A-type AFM configuration to a CAFM state takes place in the external magnetic field applied along the \textit{c} axis in the field range $3-3.5$ T. The canting angle $\theta$ decreases as a function of the external field strength and was estimated at $53^{\circ}$ at 4 T and $28^{\circ}$ at 6 T. The extrapolation to $\theta=0$ gives the field value corresponding to a full alignment of around 8.5 T, in agreement with the reported high field magnetization measurements \cite{Lai21}. Besides, our $^{55}$Mn NMR study shows a significant role of the $Mn_{Bi}$ antisites in modifying the magnetic interactions of this system. On one hand we confirm their AFM--coupling to the main Mn layers at zero field, in agreement with previous theoretical and experimental reports \cite{Sahoo24,Lai21}.  And more importantly,  we demonstrate for the first time that they introduce a new magnetic moment  to this system by providing a spin polarization of the 6s orbitals of bismuth. This moment is shown to be antiparallel to the moment of Mn antisites and thus parallel to the magnetic moment of the main Mn layers and as an additional FM component of the system may be an important consideration in engineering the QAHE devices.

\section{Acknowledgements}

We thank dr Kamil Sobczak for performing the TEM characterization of the studied sample.
R. K. was supported by The National Science Centre (NCN) grant 2022/06/X/ST3/00511.

%\cite{Novoselov04,Han14} [Novoselov2004, Han2014]  \cite{Geim13,Lee13,Xu14,Klein18,Li19} [Geim2013, Lee2013, Xu2014, Klein2018, Li2019]
%\cite{Hasan10,Qi11,Fruchart13} [Hasan2010, Qi2011, Fruchart2013]
%\cite{Liu08,Qi08,Yu10} [Liu2008, Qi2008, Yu2010] 
%\cite{Chang13} [Chang2013]
%[Sun2019, Trang2021]

\bibliography{References1}

\end{document}